\documentclass[prb,preprint,showpacs,preprintnumbers,amsmath,amssymb]{revtex4}

\usepackage{graphicx}
\usepackage{dcolumn}
\usepackage{bm}

\begin{document}


\title{A Proposed Experiment to Study Relaxation Formation of a Spherical 
Tokamak with a Plasma Center Column}

\author{S.~C.~Hsu}
\email{scotthsu@lanl.gov}
\affiliation{Physics Division, Los Alamos National Laboratory, 
Los Alamos, NM\, 87545}
\author{X.~Z.~Tang}%
\email{xtang@lanl.gov}
\affiliation{Theoretical Division, Los Alamos National Laboratory,
Los Alamos, NM\, 87545}

\date{\today}

\begin{abstract}

A spherical tokamak (ST)
with a plasma center column (PCC) can be formed via
driven magnetic relaxation of a screw
pinch plasma.  An ST-PCC could in principle eliminate many
problems associated with a material center column, a key weakness
of the ST reactor concept.
This work summarizes the design space for
an ST-PCC in terms of flux amplification, aspect ratio, and elongation,
based on the zero-$\beta$ Taylor-relaxed analysis
of Tang \& Boozer [Phys.\ Plasmas~{\bf 13},
042514 (2006)].  The paper will discuss
(1)~equilibrium and stability properties of the
ST-PCC, (2)~issues for an engineering design, and (3)~key differences
between the proposed ST-PCC and the ongoing Proto-Sphera effort in Italy.

\end{abstract}

\maketitle

\section{\label{intro} Introduction}

The spherical tokamak (ST) is a promising magnetic confinement
configuration, offering higher maximum plasma $\beta$ than conventional
tokamaks.\cite{peng86}  However, the low aspect ratio requirement results in
severe
space-constraint for the material center column, which in a
conventional ST reactor must house the inboard toroidal field (TF)
coils, the Ohmic solenoid, and shielding against fusion neutrons.  
The small center post probably eliminates the possiblity of an inboard
tritium-breeding
blanket, and it results in very high Ohmic dissipation of the
center column TF current.\cite{miller03}
An ST with a current-carrying plasma
center column (PCC) could in principle overcome many of the weaknesses
of a material center column.  The PCC used in combination
with coaxial helicity injection\cite{jarboe89} (CHI)
could serve four purposes:  (1)~ST
formation via driven relaxation of a screw pinch plasma, (2)~carry
the inboard TF coil current, (3)~help sustain the ST in steady-state
in combination with rf and neutral beams, and (4)~eliminate the need
for neutron shielding since there is no material center column to
suffer radiation damage.  For a reactor, power dissipation still remains an
issue for
a plasma center column, but the other advantages conferred are all very
compelling.

An ST-PCC can be formed via driven magnetic relaxation
of a screw pinch plasma.  The relaxation formation scheme, to be described in 
more detail in Sec.~\ref{formation},
is largely inspired from spheromak research\cite{jarboe94,bellan00}
and in particular a recent spheromak formation experiment that showed how a
current-driven kink instability of a plasma column leads to
relaxation and poloidal flux amplification.\cite{hsu03}
This approach also takes advantage
of the substantial body of work on CHI for ST startup\cite{nelson94,raman03}
and sustainment.\cite{nelson94}
Using Taylor-relaxed\cite{taylor86}
plasmas as a base-line
example, a recent theoretical and numerical study\cite{tang06}
by Tang \& Boozer
clarified the design space
of an ST-PCC in terms of
flux amplification, aspect ratio, elongation,
and vacuum bias flux.  The proposed ST-PCC is based on
the initial study by Tang \& Boozer.

This paper is organized as follows.  Section~\ref{formation} will
describe the conceptual basis for forming an ST-PCC via driven magnetic
relaxation of a screw pinch plasma.
Section~\ref{theory} will summarize the design space for an ST-PCC and discuss
the equilibrium and stability properties.
Section~\ref{design} will discuss
the primary engineering issues for designing a concept exploration (CE) class
ST-PCC experiment, and Sec.~\ref{protosphera} will discuss key
differences between the proposed ST-PCC and the Italian Proto-Sphera 
project,\cite{alladio05, papastergiou05} the first and
only other effort in the world exploring the plasma center column concept
for an ST\@.  Section~\ref{summary} provides a summary.

\section{ST-PCC formation via driven magnetic relaxation}
\label{formation}

A large body of work exists with regards to the formation of compact
toroid plasmas, such as spheromaks,\cite{jarboe94, bellan00}
via driven relaxation.  By driving current
on open bias field lines, a $\lambda$ (ratio of plasma source current to
flux) gradient is established, which
drives instabilities and causes the plasma to relax, {\em i.e.} flatten 
the spatial profile of normalized current density
($j_\|/B$).\cite{boozer86}  Thus, by driving current on
edge open field lines, relaxation will ``transport'' the edge current into
the closed-flux core region, in effect resulting in current 
drive.\cite{tang04b}
In the simplest case, a plasma column is driven kink unstable
and undergoes global relaxation, resulting in the formation of a
spheromak.\cite{sovinec01,hsu03}
The tendency of plasmas to
relax due to nonlinear processes is what makes forming spheromaks easy
and indeed almost unavoidable.  Driven relaxation formation of spheromaks
inspires and provides the foundation and a large body of
empirical results for establishing the ST-PCC concept, which will rely
on the same concept of driving a plasma column unstable and allowing it
to relax.  However, in the case of the proposed ST-PCC,
an additional CHI system will provide a new degree of freedom for
driving current-driven instabilities needed for global relaxation, and to
actively modify the $q$ profile of the relaxing PCC to ``coax'' it
into the ST state.  After ST-PCC formation, the PCC current and flux
would be adjusted to provide the ST equilibrium toroidal field.

A subtle difference between the flux-core spheromak and ST-PCC is the 
ratio between the center column flux $\chi_c$ that intercepts the electrodes
and
the closed poloidal flux $\chi_p$ in the axisymmetric $n=0$ magnetic field
after relaxation.  For a conventional gun-driven spheromak experiment,
$\chi_c/\chi_p$ is small so that the toroidal field from the plasma current
passing through the center column is small compared with the magnetic
field generated by the toroidal plasma current flowing inside the separatrix.
By increasing $\chi_c/\chi_p,$ one has the freedom to manipulate the 
toroidal field in comparison with the poloidal field, and thus the $q$ profile,
in the relaxed plasma inside the separatrix.  The chamber
geometry (elongation) has a strong effect on $q$ and provides
perhaps the most effective ``knob'' for manipulating the $q$ profile.
As will be described below, there is a
parameter regime where an ST $q$ profile can be obtained with a PCC alone,
even in the limit of complete Taylor relaxation,\cite{taylor86} {\em i.e.},
a spatially uniform $\lambda$ profile.
The proposed 
ST-PCC formation scheme will be based on this unusual relaxation solution
branch.

Controlled decay of spheromaks has resulted in several-hundred eV 
electrons,\cite{Jarboe90,mclean06}
implying the existence of nested closed magnetic flux surfaces.
This fact provides promise that a CE-class
experiment will be able to obtain $T_{\rm e} \approx 100$--$200$~eV ST-PCC
plasmas with good particle confinement.  This target plasma can then be
sustained in follow-on research
using other non-inductive current drive and heating schemes, such
as rf and neutral beam injection.

The sustainment phase of an ST-PCC requires consideration of
additional physics issues.
First, the relaxed plasma after the initial formation stage typically has low 
$\beta$ and a rather flat $j_\parallel/B$ profile, both of which reduce
the energy drive for MHD instabilities.
The sustainment aims to ramp up the plasma pressure towards a conventional
high-$\beta$ equilibrium. The effect of a PCC (instead
of a material center column) on the stability of external MHD modes must be
understood.  Second, an electrostatic bias must be present thoughout the
discharge to maintain the toroidal magnetic field in an ST-PCC plasma.
How the externally imposed electric field affects transport properties via
${\bf E}\times{\bf B}$ flow\cite{tang03,tang04a}
must also be understood in order to assess the
attractiveness of the concept.

\section{ST-PCC design space}
\label{theory}

A recent theoretical and numerical study\cite{tang06} by Tang \& Boozer
based on zero-$\beta$
magnetically relaxed (spatially uniform $\lambda$) equilibria has clarified
the design space for an ST-PCC\@.  Although a real
ST-PCC will have finite $\beta$ and non-uniform $\lambda$ profiles,
a study based on Taylor-relaxed plasmas was used as a base-line
example to provide insight into the equilibrium properties of a
driven relaxation
experiment in a simply-connected chamber.  (Of course, finite $\beta$
equilibrium and stability studies 
and nonlinear MHD simulations must be performed to justify and support
the design of a real experiment.)  The zero-$\beta$ study showed that 
two essential factors influence the ST-PCC design space:
(1)~flux amplification is
directly related to the aspect ratio (ratio of major to minor radius) and
(2)~plasma elongation $\varepsilon$ (ratio
of chamber height to radius)
has a large affect and vacuum bias flux $\chi_0$ a smaller subtle effect
on the $q$ profile.

Tang \& Boozer numerically solved the force-free equation
$\nabla \times \mathbf{B} = \lambda \mathbf{B}$ for a finite length
cylinder and $\chi_0$ on the boundary.  They showed that relaxed
ST-PCC equilibria exist in a particular parameter regime
described below.
They investigated the relationship between aspect ratio and
flux amplification and found that the aspect ratio scales approximately
inversely proportional to the flux amplification factor, independent
of $\varepsilon$ (see Fig.~3 of
Ref.~10).  An
ST-relevant aspect ratio of 1.5--2 is achieved with a flux amplification
factor of 1--2.5, which is routinely achieved in laboratory
helicity injection experiments.  A smaller aspect ratio down to 1.25 would
require a flux amplification factor of 9, which may be achievable but
probably not a good choice for a conservative ST-PCC experimental design.
They also investigated the
$q$ profile as a function of $\varepsilon$ and showed that $q_{\rm edge}$
scales approximately linearly with $\varepsilon$ from $\varepsilon \approx
0.5$--3
(see Fig.~4 of Ref.~10).
To achieve $q>1$ throughout the interior of the separatrix, an
$\varepsilon = 2$--3 is
required (see Fig.~6 of Ref.~10).

Thus, high $\varepsilon$ and high flux amplification are desirable for
forming an ST-PCC equilibrium in order to achieve $q>1$ and
low aspect ratio, respectively, as required for an ST\@.
ST-PCC stability, however, constrains both $\varepsilon$ and flux amplification.
These constraints can be understood through the properties of the
force-free (Chandrasekhar-Kendall\cite{chandrasekhar57}) eigenmode solutions
of the discharge chamber, which entirely determines the stability
for a relaxed plasma.
For $\varepsilon\lesssim 1.67$, the stability of the lowest
energy axisymmetric relaxed state $\lambda^{(n=0)}_1$
is guaranteed because it is the overall
lowest energy state of the system.  For $\varepsilon \gtrsim 1.67$, the
lowest energy helical eigenmode $\lambda^{(n=1)}_1$
becomes a lower energy state than $\lambda^{(n=0)}_1$.\cite{finn81,
bondeson81}
This means that if one forms an axisymmetric
relaxed state for $\varepsilon > 1.67$ at $\lambda > \lambda^{(n=1)}_1$, the
plasma will be unstable.  Thus, for $\varepsilon \approx 2$--3 as required for
$q>1$, the ST-PCC $\lambda$ must be less than $\lambda^{(n=1)}_1$ for stability.
The latter limits flux amplification which scales as
$[(\lambda^{(n=0)}_1)^2 - \lambda^2]^{-1}$.\cite{tang06}  It should
be noted that the vacuum bias flux can be adjusted to optimize
the flux amplification at a given $\lambda$.\cite{tang06}

Table~\ref{table} summarizes the design space for an ST-PCC\@.
For a much more 
detailed discussion of relaxed ST-PCC equilibria and stability, the
reader is referred to Ref.~10.  Further studies are necessary to
characterize the equilibrium and assess the stability of finite-$\beta$ driven
ST-PCC's, and for designing an optimized experiment.

\section{Engineering design issues}
\label{design}

\begin{figure}
\includegraphics[width=6truein]{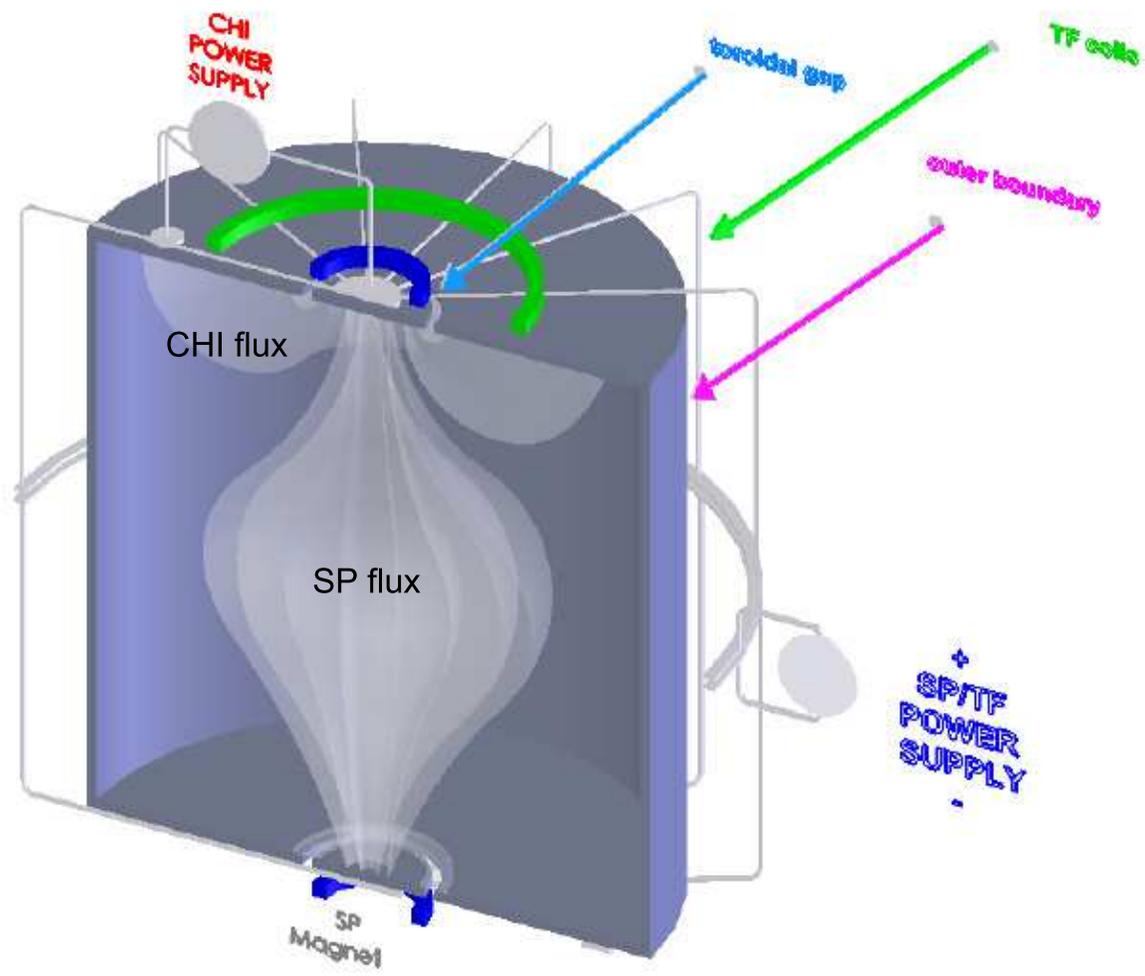}
\caption{\label{setup} Conceptual drawing of proposed ST-PCC experiment,
identifying key components:  flux-conserving boundary, screw pinch (SP)
and TF
circuit and power supply, SP magnets for SP axial flux, CHI magnet
for CHI vacuum bias flux, CHI power supply,
toroidal insulating gaps, and SP cathode/anode inside the toroidal
gaps.}
\end{figure}

An experiment to form an ST-PCC via driven relaxation of a screw
pinch plasma would benefit from having two sets of biased electrodes
with a cylindrical boundary.  A primary consideration is to
have the versatility to drive a $\lambda_{\rm PCC}$ that is different
from the relaxed $\lambda_{\rm ST}$.  This provides freedom to drive
relaxation by exciting either a center column kink or an open flux kink
on the CHI bias flux.  A further benefit of a separate CHI system
is the natural formation of a single null plasma where edge
current can be tuned independently of the center column current.  This
would become especially important for ST-PCC sustainment.
Figure~\ref{setup} shows a conceptual
drawing for a CE ST-PCC experiment,
including all the important sub-systems.  Below, each
sub-system will be discussed separately along with their primary issues.

{\em Vacuum chamber and boundary:}  Shape and geometry ($\varepsilon \approx
2$--3) are chosen to
optimize coupling between PCC and ST during driven relaxation formation,
as well as equilibrium and stability of the formed ST-PCC\@.  Wall thickness
must allow bias flux penetration while acting as a flux-conserver for
the plasma.  The boundary requires electrical breaks
that will allow the application of bias voltages for driving PCC and CHI
current.  Toroidal gaps in the boundary on either end of the geometric axis
of the device define the electrical breaks for the PCC cathode and anode.
The gaps also allow voltage to be applied between the cathode and
the outer boundary, as well as between the anode and the outer boundary.
Together with the CHI flux, this injects magnetic helicity into the
system.  The shape of the flux conserving boundary
would be chosen to
optimize equilibrium and stability properties of the ST-PCC, and to
facilitate engineering design of the couplings between PF, TF, and PCC power
systems.

{\em Screw pinch PCC:}  The PCC
would be created between electrodes located at either end of the geometric axis
of the
device.  Via current ramp-up, the
PCC would undergo driven relaxation in concert with the
CHI system to form an ST and subsequently carry equilibrium TF coil current.  
The power system for this plasma source would be a
high voltage 
capacitor bank to create the PCC 
and to sustain the PCC current below pinch instability thresholds
for
generating steady TF after relaxation formation.  
A CE experiment would likely need to last only
several milliseconds, which is enough time for a plasma of a few tens of
eV to globally relax and settle into the ST-PCC equilibrium.
An arc discharge between tungsten sprayed copper electrodes
would be used, similar to spheromak and other
helicity injection experiments.  The current (several hundred kA)
and power (tens of MW) would be comparable to SSPX.\cite{hill03}
In a CE experiment, the power dissipation of the center column is probably
not an issue.  However, this is a key issue when considering how an ST-PCC
would scale to a reactor.  Even if the PCC could be maintained at an
elevated temperature of a few hundred eV, the power dissipation in the PCC for a
1~GW reactor would probably be on the order of 100~MW\@.  This issue needs
careful consideration and study for the reactor viability of the ST-PCC concept.

{\em TF coils:}  A key engineering
issue is coupling the outboard TF coil windings to the PCC, which
closes the TF coil circuit.  In conventional tokamaks and ST's, the TF coil
is a single solenoidal winding with
$n$ turns.  With a PCC, the TF coil system would actually be $n$ turns each
connected in parallel, with the PCC being shared by every parallel turn.
For a CE experiment, the power source for the TF coils would be the same PCC
capacitor bank.

{\em CHI system:}  The CHI 
system applies voltage
across the inner and outer boundaries, which is linked by a
bias magnetic flux. 
The CHI bias flux would be created by a combination of the CHI and PCC bias
flux coils.  The CHI current is expected to be on the order
of several tens of kA driven at a few hundred Volts.

{\em PCC and CHI bias flux coils:}  PF coil sets
are needed to provide the poloidal fluxes needed to form and
sustain an ST-PCC:  (1)~axial flux for the PCC and (2)~bias flux linking the
inner and outer boundaries for CHI\@.  The two fluxes would need independent
timing and control and thus would require independent power sources.
A CE experiment would likely
require several to several tens of mWb for each system.

{\em Diagnostics:}  The primary diagnostics needed for a CE experiment
are arrays of magnetic probes to characterize ST-PCC formation and 
equilibria, as well as voltage and current diagnostics to characterize
the PCC and CHI sources.  Magnetic (Mirnov) coils would be placed
at different toroidal locations on the vacuum chamber
wall to monitor toroidal mode activity.  Fast camera imaging of global
plasma light emission would provide information on plasma evolution
and guide the placement of quantitative probe diagnostics.  Simple electrostatic
probes, such as Langmuir probes to measure $n_{\rm e}$ and $T_{\rm e}$ and
Mach probes to measure ion flow, would also be used.
In later stages of a CE experimental program, when
higher temperatures in the hundreds of eV range are expected, advanced
diagnostics would be proposed and implemented, {\em e.g.}, Thomson scattering
to measure $T_{\rm e}$ and interferometry to measure $n_{\rm e}$.

\section{Comparison with Proto-Sphera}
\label{protosphera}

The Italian Proto-Sphera project\cite{alladio05,papastergiou05}
is under construction and is the first project to
study the ST-PCC concept.  However, there are several key differences
between the present proposed ST-PCC and Proto-Sphera.
It would be of interest to explore the relative merits of the two approaches
with complementary experimental programs.

Proto-Sphera will use an emissive electrode system to form the
PCC\@.  This system is potentially capable of high power handling,
relevant for eventual use on proof-of-principle and proof-of-performance
class experiments.  In contrast, the proposed ST-PCC 
would use a much simpler arc discharge system at short pulse length to test the
relaxation formation scheme.

The proposed ST-PCC formation scheme and target
equilibrium are fundamentally different from those of Proto-Sphera,
which proposes
to use induction via a quick swing of the poloidal field coil current
followed by compression of the plasma to raise the $q$ 
profile.\cite{alladio05}  The Proto-Sphera
equilibrium was originally motivated by higher order Chandrasekhar-Kendall
modes, and the reference Proto-Sphera equilibrium removes two higher order
lobes while retaining a high $\varepsilon$ center lobe and PCC.\cite{rogier03}
In contrast, the proposed
ST-PCC is based on relaxed states similar to a flux-core spheromak, and
would be formed via driven relaxation (likely kink instability) of the
PCC and possibly CHI flux.

The final key difference is the initial
sustainment method after formation.  Proto-Sphera will
use helicity injection via the PCC itself,
while the proposed ST-PCC uses an independent CHI system which would
also play an integral role in the formation phase.

\section{Summary}
\label{summary}

A plasma center column would eliminate several key weaknesses of the
ST reactor concept, although PCC power dissipation in a reactor scale
experiment could remain a concern requiring further study.
An ST-PCC can be formed
via driven magnetic relaxation of a screw pinch plasma.  The idea
of driving a plasma column kink unstable and allowing the ensuing
global relaxation to provide the final desired ST-PCC equilibrium is
inspired from spheromak research and helicity injection
via magnetized coaxial guns.
An initial theoretical and numerical study by Tang \& Boozer\cite{tang06}
has clarified the design space for zero-$\beta$ Taylor-relaxed ST-PCC's.
They showed that for high elongation of 2--3 and modest flux
amplification factors of 1--2.5, stable relaxed ST-PCC equilibria with $q>1$
and aspect ratios of 1.5--2 exist.  These results are promising and
motivate further studies of finite-$\beta$ ST-PCC equilibrium and stability
leading to an optimized experimental design.
A representative ST-PCC experiment would feature two sets of biased
electrodes.  One set would form the PCC and drive it unstable for relaxation.
The other set would be a CHI system, which would provide additional 
freedoms in relaxation formation and raising the $q$ profile after
the initial relaxation.  A concept exploration ST-PCC experiment would be
comparable
to SSPX in both its hardware and plasma parameters, although the 
initial research emphasis would be on the relaxation formation scheme.

An ST-PCC concept exploration experiment is needed for detailed
studies of the basic plasma physics issues for relaxation formation
and non-inductive sustainment, as well as many of the engineering issues
which will ultimately determine the reactor potential of the ST-PCC concept.

\acknowledgments{The authors thank Mr.~Mark Kostora for the conceptual
design drawing.  This work was supported by the U.S. Department of
Energy Office
of Fusion Energy Sciences under contract no.~W-7405-ENG-36, the
Los Alamos Laboratory Directed Research and Development (LDRD) Program,
and a Los Alamos Frederick Reines Fellowship (SCH).}


\newpage

\begin{table}[t]
\centering
\begin{tabular}{l l} \hline \hline
ST-PCC requirement \hspace{.2truein} & Impact on design (target)\\ \hline
$\langle q \rangle_{\rm separatrix} > 1$ & $\varepsilon > 2$ (2--3) \\
aspect ratio 1.5--2 & determined by flux amplification (1--2.5) \\
stability & lower $\varepsilon$, lower flux amplification\\
efficiency & higher flux amplification\\
\hline \hline
\end{tabular}
\caption{Summary of ST-PCC design space, showing some competing factors
between equilibrium and stability.  A reference design would have
$\varepsilon \approx 2$--3 and a modest flux amplification factor
of 1--2.5.}
\label{table}
\end{table}

\end{document}